# Deviants' Dilemma – The Simulation[*]

**W. A. T. Wan Abdullah[1]**
*Physics Department*
*Universiti Malaya*
*50603 Kuala Lumpur, Malaysia*

**A. K. M. Azhar[2]**
*Graduate School of Management*
*Universiti Putra Malaysia*
*43400 Serdang, Malaysia*

**Abstract.** The Deviants' Dilemma is a two-person game with the individual gain conflicting with the choice for global good. Evolutionary considerations yield fixed point attractors, with the pehenomena of exclusion potentially playing an important role when current opponent information is available. We carry out computer simulations which substantiates and illuminates theoretical claims, and brings to light the pertinance of the choice between deterministic and stochastic dynamics, and the conjecture of 'ergodicity spread'.

**Introduction**

In social and economic systems, agents interact with each other through games or localized potentials, in the sense that different agents are subjected to different pertaining potentials, which are dependent on the states (or moves) of the other agents, thus allowing the dislocation of individual optima from the global optimum. Interesting cases are when the agents' respective optimum states ("Nash state") do not constitute the community's optimum ("Pareto state"), as given by the Prisoner's Dilemma and its multiplayer version the Public Goods Game. This has become the paradigm for studying the evolution of Paretoic cooperativity from selfish Nash tendencies [Axelrod, 1984].

Recently, the Deviants' Dilemma, which falls into the same category of games, has been proposed [Azhar et al., 2003] to model the run-up to seniority in institutions [Frank, 1989]. A theoretical evolutionary analysis of the game was carried out; it remains for this paper to provide computer simulation studies to test the theoretical predictions and to further elucidate the fine ramifications that may lie hidden in pure conceptualization. Furthermore, when information on current opponent states are available (albeit at a cost), there is the potential that a Nash state may loose out by being excluded as Pareto pairs are formed.

We describe briefly the Deviants' Dilemma in the next section, and the simulation studies and results in those following it.

---

## Deviants' Dilemma

The Deviants' Dilemma (DD) game is a two-person interaction game with the following symmetric 'fixed zero baseline' 2×2 payoff matrix:

| A \ B    | Pseudo | Deviant |
|----------|--------|---------|
| Pseudo   | a / a  | a / b... |

| B \\ A  | Pseudo | Deviant |
|---------|--------|---------|
| Pseudo  | a, a   | b, 0    |
| Deviant | 0, b   | c, c    |

with the requirement that $0 < a < c < b$. The evolutionary stable state (ESS) analysis yields that the fraction of pseudos, $p$, tends to 1.0 for all initial states, given rational choice. If the agents are allowed to check on the state of their prospective partners with the cost $r$, then $p$ has a fixed point of $1-(r/c)$.

## Simulation

We carried out computer simulations of a population of 1000 agents playing the DD game with random partners, at least for 50 timesteps (plays per agent). We chose $a=2$, $b=6$ and $c=4$ for the payoffs. We utilized for comparison two different updating schemes – deterministic (i.e. rational agents), where moves are played depending on direct comparison for better score between between a pseudo and being a deviant, and probabilistic, where moves are chosen randomly depending on the ratio of the scores. We also tried two methods for choice of play: one where the moves are based on the average of every agents immediate previous scores for the two states, and one where the moves are based on the average scores from only the individual agents' own histories (but of all previous scores).

When checking on partners is chosen, agents check prospective partners until advantageous ones are found or until none of such are found when they are forced to play with whosoever they found last. A cost of 1 to the payoff is incurred when an agent checks.

## Results and Discussions

The results for deterministic updates, without partner-checking, with different initial pseudo fractions, are given in Figure 1 for moves based on everyone's previous scores, and in Figure 2 for moves based on own histories. Figure 1 supports ESS analysis, while more 'myopic' agents pose some problems in Figure 2. In fact, the latter is worse if we start the agents with zero memory, and any initial move can freeze state choices to the limited initial set, and thus seriously breaking ergodicity, especially with probabilistic updating. (In our simulations we have put some arbitrary initial scores for both states in the agents' histories.) A kind of 'ergodicity spread' conjecture, where the representative statistical ensemble is reflected both horizontally in the agent population score distribution and vertically in the individual agent history, seems unjustified.

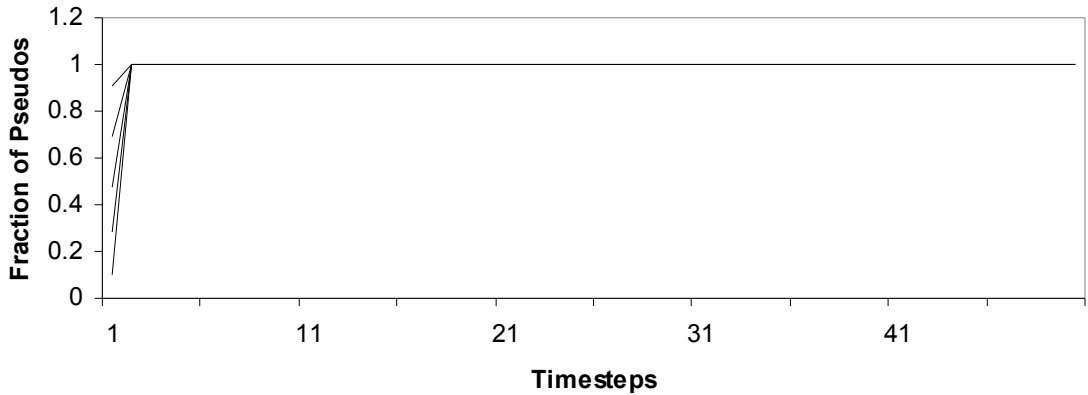

FIGURE 1

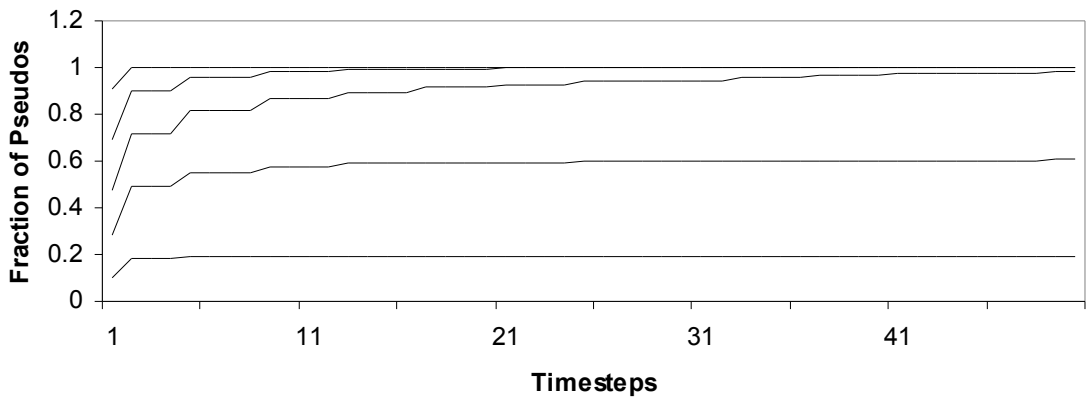

FIGURE 2

Figures 3 and 4 show the evolution with probabilitic updating for moves based on everyone's previous score and one's own history respectively. Here a behaviour different from that with rational choice is seen. For both cases, there appears to be an attractor around $p \sim 0.7$. In the second case, for initially many pseudos, there is an initial dip in pseudos before the ratio levels up at the attractor. Probabilistic updating incorporates some kind of noise, which allows more 'exploration' of the phase space, and thus better ergodicity, as compared to deterministic updating. The seeming appearance of the fixed point is not understood as yet and requires further scrutiny.

**Probabilistic Moves based on Everyone's Previous Score**

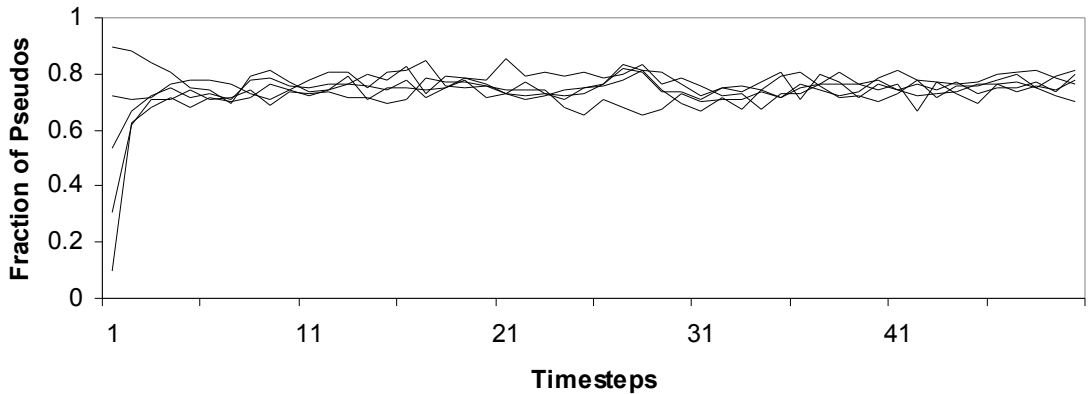

FIGURE 3

**Probabilistic Moves based on Own History**

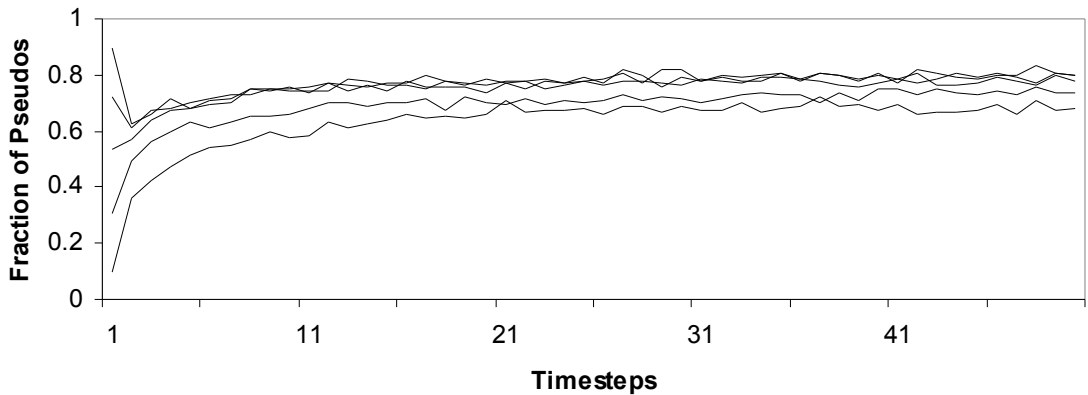

FIGURE 4

Partner checking seems not to change signifcantly the results obtained, except for the case of probabilistic moves based on the average of everyone's previous score. This is shown in Figure 5. Now 'pseudomania' dominates fully, and there is no sign of the expected fixed point at $p = 0.75$ (except for the case of probabilistic moves based on own history, but then we need to explain how the fixed point is there even without partner checking, where pseudos are expected to dominate fully). The expected effect of exclusion is not obvious, possibly because the time to system stability is too small for agents to have time learn to make Pareto partners.

Dynamics of computer simulations may bring about subtle differences to the theoretical ESS analysis where some hidden assumptions may have been made. This could lead to the differences in the expected and obtained results. For instance, when an agent choses to check on his partner, theoretical analysis assumes that the agent knows what gain he would have by doing so. In the simulations, this has got to be represented in the scores available upon which the agent makes his choice, and it is not obvious that the scores, whether form everyone's previous experience, or from own histories, *should* contain this information (intrinsic in the scores), or at least *could* contain this information (i.e. enough ergodicity).

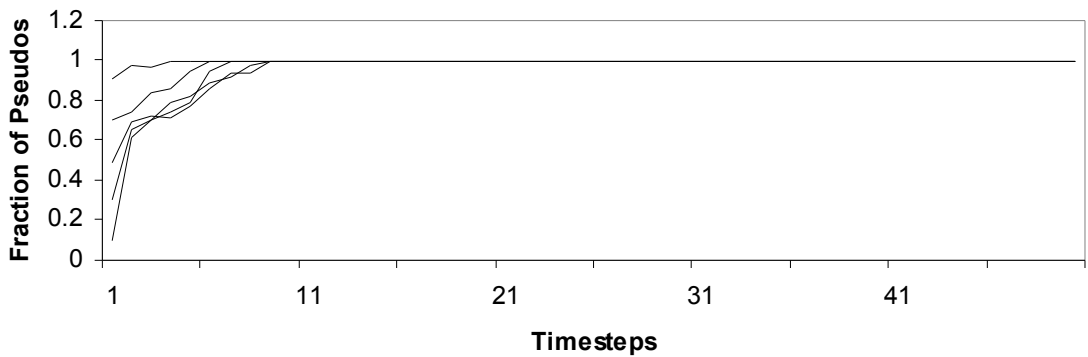

**Probabilitic Moves based on Everyone's Previous Score, Checking Partner**

FIGURE 5

**Conclusions**

In the simulation studies of the Deviants' Dilemma, we show agreement of the ESS analysis for pseudo dominance for agents with rational choice and looking at other agents' performances. However, questions are raised from the different behaviour of systems with probabilistic moves and with myopic agents, while further scrutiny is required for systems with partner-checking, to bring out subtleties in the dynamics which the ESS analysis might have overlooked.

The questions and discrepancies are perhaps the reflection of the non-fully ergodic nature of the system, due to the limited time to sytem stability. Can this in turn be a characteristic of social and economic systems which differentiates them from statistical physical ones?

**Acknowledgements.** AKM Azhar acknowledges the support of an MPKSN IRPA grant.

**References**

Axelrod, R (1984) *The Evolution of Cooperation*. Basic Books, New York.

Azhar, AKM, Wan Abdullah, WAT, Kidam, ZA, and Salleh, A (2003) *The Deviants' Dilemma: A Simple Game in the Run-Up to Seniority.* Presented at WEHIA 2003, Keele; arXiv e-print cond-mat//0310340.

Frank, Robert (1989) *Passions within Reason: The Strategic Role of the Emotions.* W W Norton, New York.